\documentclass[preprint,amsmath,amssymb,aps]{revtex4-2}
\usepackage{graphicx}
\usepackage{dcolumn}
\usepackage{bm}
\usepackage{xcolor}
\usepackage{booktabs}

\newcommand\zfig[1]{{\color{blue}#1}}

\begin{document}

\title{Disentangling Electronic and Phononic Thermal Transport Across 2D Interfaces}

\author{Linxin Zhai, Zhiping Xu}
\email{xuzp@tsinghua.edu.cn (Z.X.)}

\affiliation{Applied Mechanics Laboratory, Department of Engineering Mechanics, Tsinghua University, Beijing 100084, China}

\date{\today}

\begin{abstract}
Electrical and thermal transport across material interfaces is key for 2D electronics in semiconductor technology, yet their relationship remains largely unknown.
We report a theoretical proposal to separate electronic and phononic contributions to thermal conductance at 2D interfaces, which is validated by non-equilibrium Green's function calculations and molecular dynamics simulations for graphene-gold contacts.
Our results reveal that while metal-graphene interfaces are transparent for both electrons and phonons, non-covalent graphene interfaces block electronic tunneling beyond two layers but not phonon transport.
This suggests that the Wiedemann-Franz law can be experimentally tested by measuring transport across interfaces with varying graphene layers.
\end{abstract}

%

\maketitle
\newpage

With the rise of nanoelectronics in the `more-than-Moore' technology, two-dimensional (2D) materials show great potential due to their unconventional electronic structures and transport characteristics.
Their single- or few-atom-layer thickness allows for effective contact or gating with extremely thin and short channels~\cite{das2021transistors,wu2022vertical}.
Charges in graphene carry zero effective masses and exhibit ballistic transport, resulting in an ultra-high electrical conductivity up to $10^{6}$ S/m and a current density capability six order-of-magnitude higher than that of copper~\cite{geim2009graphene,lim2021measurements}.
The diverse chemistry of transition metal dichalcogenides (TMDs) provides a wide range of electronic properties, from semiconducting to metallic, and even superconducting~\cite{song2017progress}.
Black phosphorus exhibits high carrier mobility and a widely tunable bandgap, which can be adjusted through the number of layers and the application of an external field~\cite{xia2019black}.
Combining the advantages of 2D materials in their layered structures opens up new opportunities for device applications~\cite{geim2013van}.

For instance, concepts such as floating and Moir{\'e} gating have been introduced to develop neuromorphic functionalities, including memristors~\cite{migliato2023large} and synaptic transistors~\cite{yan2023moire}.
The strength of interfacial electrical and thermal coupling between 2D materials and substrates significantly impacts charge carrier mobility and heat dissipation in 2D devices~\cite{ye2023ultra,wang2014critical,wang2016intercalated}.
In 2D devices such as field-effect transistors (FETs), material pairing at contacts between metallic electrodes and 2D materials defines their Schottky or Ohmic nature and the rectifying characteristics~\cite{liu2018approaching,kwon2022interaction}.
The contact can be further engineered by, for example, interaction or pressure across the interface~\cite{wang2016water,bian2023vertical}.

The diminishing thickness of 2D structures enables not only efficient gating and control~\cite{das2021transistors} but also effective electronic and thermal coupling across 2D interfaces, which can be leveraged to control device performance.
However, the power capacity of highly integrated 2D nanoelectronics is limited by the need for effective thermal management to ensure efficient power dissipation~\cite{avouris2010graphene, xu2012heat, yalon2017energy}.
In this context, energy transfer processes across 2D interfaces are critical.
Understanding electronic transport and heat dissipation across these interfaces is essential for enhancing the performance of current devices and for developing new designs.

The Wiedemann-Franz (WF) law~\cite{jones1985theoretical} offers a theoretical framework to understand the coupling between electronic charge and heat transport processes.
The law states posits that the thermal ($\kappa_{\rm e}$) and electrical conductivities ($\sigma$) carried by electrons in a metal are related by a numerical factor that increases with temperature $T$, $\kappa_{\rm e}/\sigma= L_{0}T$.
Following the assumption that electrons do not interact with each other and form a degenerate Fermi-Dirac assembly, the
scattering of the electrons is due to impurities or lattice vibrations and is elastic, the Lorenz number has been showed to be a material-independent constant, $L_{0} = 1/3\pi^{2}\left(k_{\rm B}/e\right)^{2} = 2.44\times 10^{-8}$ W$\Omega \rm{K}^{-2}$.
The WF law has been validated for numerous materials and extended to apply to their interfaces~\cite{chester1961law,mahan1999wiedemann,wilson2012experimental,bürkle2018probe}.
Comparisons of the interfacial thermal conductance (ITCs) and electronic resistance of Pd/Ir interfaces at $4$ K support the interfacial WF law~\cite{wilson2012experimental,acharyya2009specific}, showing that the measured Lorenz number is consistent with $L_{0}$ within approximately $10\%$.
The law has been further extended to metal-semiconductor interfaces~\cite{bartkowiak2001heat} and molecular systems~\cite{craven2020wiedemann}.
However, this speculation across 2D interfaces has not been assessed.

In this work, we study electronic and phononic thermal conductance across Au$\lvert$$n$-layer graphene ($n$G)$\rvert$Au contacts.
An unusual layer-number dependence of the transport characteristics is revealed, which disentangles the contributions of electrons and phonons to thermal conductance and can be deployed to assess the validity of the interfacial WF law.


At the face-to-face contact between metals and 2D materials, several modes of electronic transport may be activated, including thermionic emission and quantum tunneling.
Ballistic quantum transport is often assumed in pristine 2D devices with negligible scattering from the defects and under weak electron-phonon scattering, where electrons propagate elastically in the crystal lattice~\cite{datta2005quantum}.
The conductance of a device in contact with two electrodes is then defined by the number of conducting modes $N$ and calculated from the uncertainty principle as $G = NG_{0} = 2Ne^{2}/h$, where $G_{0}$ is the fundamental conductance, $e$ is the electron charge, and $h$ is the Planck constant.

Landauer and B\"uttiker~\cite{datta2005quantum} pointed out that $G$ can be calculated from the transmission coefficients ($t_{\alpha\beta}$) between different transport modes or channels ($\alpha, \beta$) as $G = 2e^{2}/h\sum_{\alpha,\beta} \left | t_{\alpha\beta} \right | = G_{0}{\rm Tr}{\bf T}$.
The transmission matrix (${\bf T}$) and transmission $t = {\rm Tr}{\bf T}$ are calculated in this work using the non-equilibrium Green’s function (NEGF) formalism~\cite{datta2005quantum} in the framework of density functional theory (DFT) implemented with numerical atomic orbitals (NAOs)~\cite{garcia2020siesta}.
The transmission matrix $\mathbf{T}$ can be decomposed into its eigenvalues and eigenvectors, where each eigenvalue represents the transmission probability of an eigenchannel.
The conductance mechanisms can then be analyzed by identifying which quantum states are responsible for electron transport~\cite{paulsson2007transmission}.

The generalized gradient approximation (GGA) in the Perdew-Burke-Ernzerhof (PBE) parametrization~\cite{perdew1996generalized} is chosen for the exchange-correlation functional, and double-$\zeta$ polarized (DZP) orbitals as the local basis sets.
An energy cutoff of $75$ Hartree is used for the density mesh.
A $9\times9\times1$ Monkhorst-Pack grid is used for ${\bf k}$-point sampling.
The choices of these parameters are validated by yielding an energy convergence below $1$ meV/atom in the self-consistent field (SCF) calculations.
A $40\times40\times131$ ${\bf k}$-point mesh is used in calculating the transmission coefficients, reaching a convergence threshold well below the values of our concern.

\begin{figure*}[htbp]
\centering
\includegraphics[scale = 0.8] {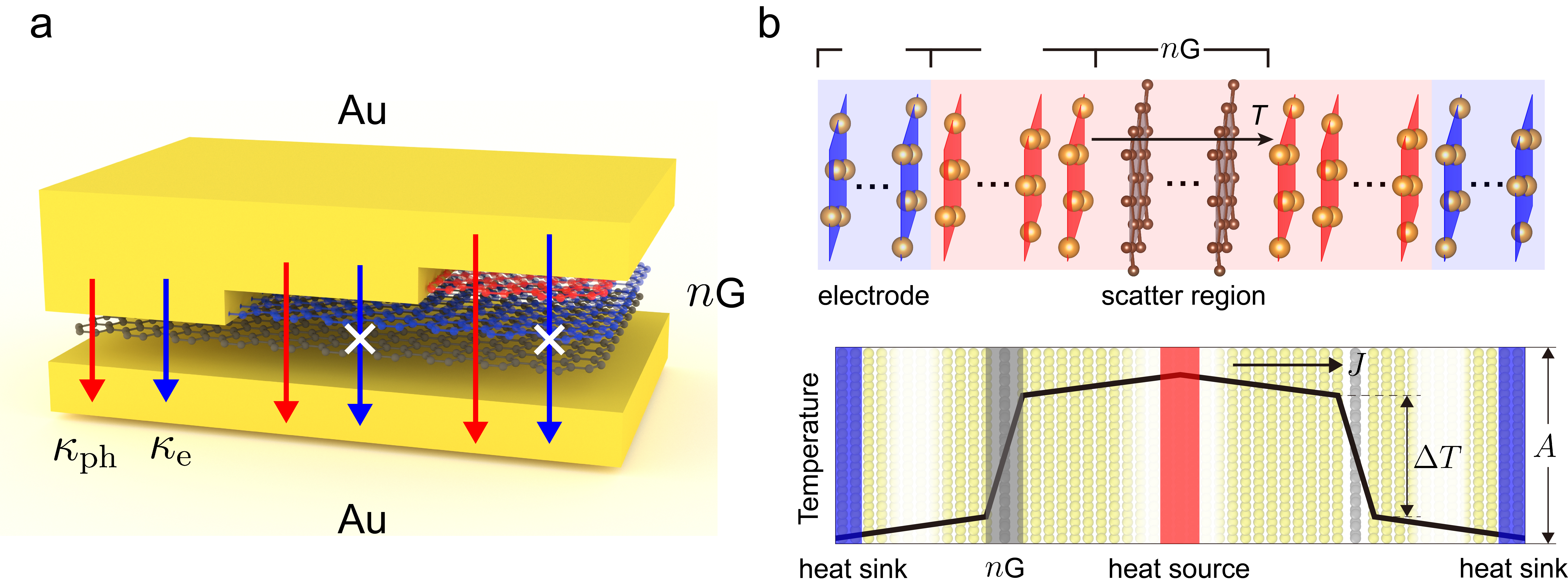}
\caption{\textbf{Energy transport across a 2D junction.}
\textbf{a.} Schematic diagram of a Au$\lvert n$G$\lvert$Au junction ($n = 1, 2, 3$).
\textbf{b.} Atomic-level structures of the electron transport setup in non-equilibrium Green's function (NEGF) calculations.
\textbf{c.} Setup and typical temperature profiles in non-equilibrium molecular dynamics (NEMD) for phononic thermal transport.
}
\label{fig1}
\end{figure*}

In the sandwich Au$\lvert n\mathrm{G}\lvert$Au device setup, the central region consists of twelve $\left( 111 \right)$ Au layers with $\sqrt{3} \times \sqrt{3} = 3$ atoms in each, or a $2 \times 2$ supercell of graphene (\zfig{Fig. \ref{fig1}a}).
The lattice constant of face-centered cubic Au is $4.169$ \AA, and the electrodes contain $3$ layers of them.
The graphene sheet is strained biaxially by $\sim 0.98\%$ to match the Au lattice, where one of the C sublattices sits on top of the Au atoms.
Periodic boundary conditions (PBCs) are enforced in the lateral directions.
The atomic-level structures are relaxed to reach a force-on-atom threshold of $0.01$ eV/\AA.
The Au-graphene and graphene-graphene distances in optimized Au$\lvert n$G$\lvert$Au ($n = 1,2,3$) geometries are summarized in \zfig{Table 1}.

\begin{table}[htbp]
  \centering
  \caption{Au-graphene (G) and G-G distances in optimized Au$\lvert n$G$\lvert$Au ($n$ = $1$, $2$, $3$) geometries.}
  \label{tab:mytable}
  \begin{tabular}{lllll}
  \toprule
  Structure & $d_{\rm Au-G}$ (\AA)  & $d_{\rm G-G}$ (\AA)\\
  \midrule
  Au$\lvert 1{\rm G} \lvert$Au & $3.30$ \AA & - \\
  Au$\lvert 2{\rm G} \lvert$Au & $3.33$ \AA & $3.48$ \AA \\
  Au$\lvert 3{\rm G} \lvert$Au & $3.32$ \AA & $3.46$ \AA \\
  \bottomrule
  \end{tabular}
\end{table}

The phononic contribution to ITC is calculated by non-equilibrium molecular dynamics (NEMD) simulations, using the large-scale atomic/molecular massively parallel simulator (LAMMPS)~\cite{plimpton1995fast}.
The time step to integrate Newtonian equations of motion is $0.5$ fs, ensuring energy conservation and yields converged predictions of ITCs.
The interatomic interactions between Au atoms are modeled using the embedded-atom method (EAM)~\cite{sutton1990long}, while graphene is described using the optimized Tersoff potential~\cite{lindsay2010optimized}. 
Non-covalent interactions between carbon atoms in different graphene layers and between carbon and gold atoms are modeled using $12-6$ Lennard-Jones potentials.
The parameters for these potentials are fitted to the binding distances and energies ($E_{\rm b}$) obtained from DFT calculations, using the same settings as those in our study of electron transport (\zfig{Table 2}).

\begin{table}[htbp]
  \centering
  \caption{$12-6$ Lennard-Jones parameters and binding energies for interfacial interactions.}
  \label{tab:mytable}
  \begin{tabular}{lllll}
  \toprule
  Interface & $\sigma$ (\AA)  & $\epsilon$ (meV)  & $E_{\rm b}$ (meV/\AA$^{2}$)\\
  \midrule
  C-C & 3.45  & 1.77 & 11.3\\
  C-Au & 3.34  & 7.54 & 19.1\\
  \bottomrule
  \end{tabular}
\end{table}

To construct the Au$\lvert n{\rm G}\lvert$Au junctions, we expand the models in electron transport studies into an $8\times8$ supercell in the lateral directions.
Each electrode is extended to $20$ nm along the transport direction, and the structure is mirrored into the M\"uller-Plathe setup~\cite{muller1997simple}.
The composite system is equilibrated using a Nos\'e-Hoover thermostat at $300$ K for $250$ ps.
Elastic collisions between the hottest atom in the heat source and the coldest atom in the heat sink are forced every $150$ fs.
This process of kinetic energy exchange drives a temperature difference $\Delta T$ and a thermal flux $J$ across the cross-section area $A$.
All NEMD simulations run for $4$ ns to reach the steady state, where a spatial temperature profile is extracted.
The ITC is then calculated as $\kappa = J/\left(\Delta TA\right)$.

\begin{figure*}[htbp]
\centering
\includegraphics[scale = 0.8] {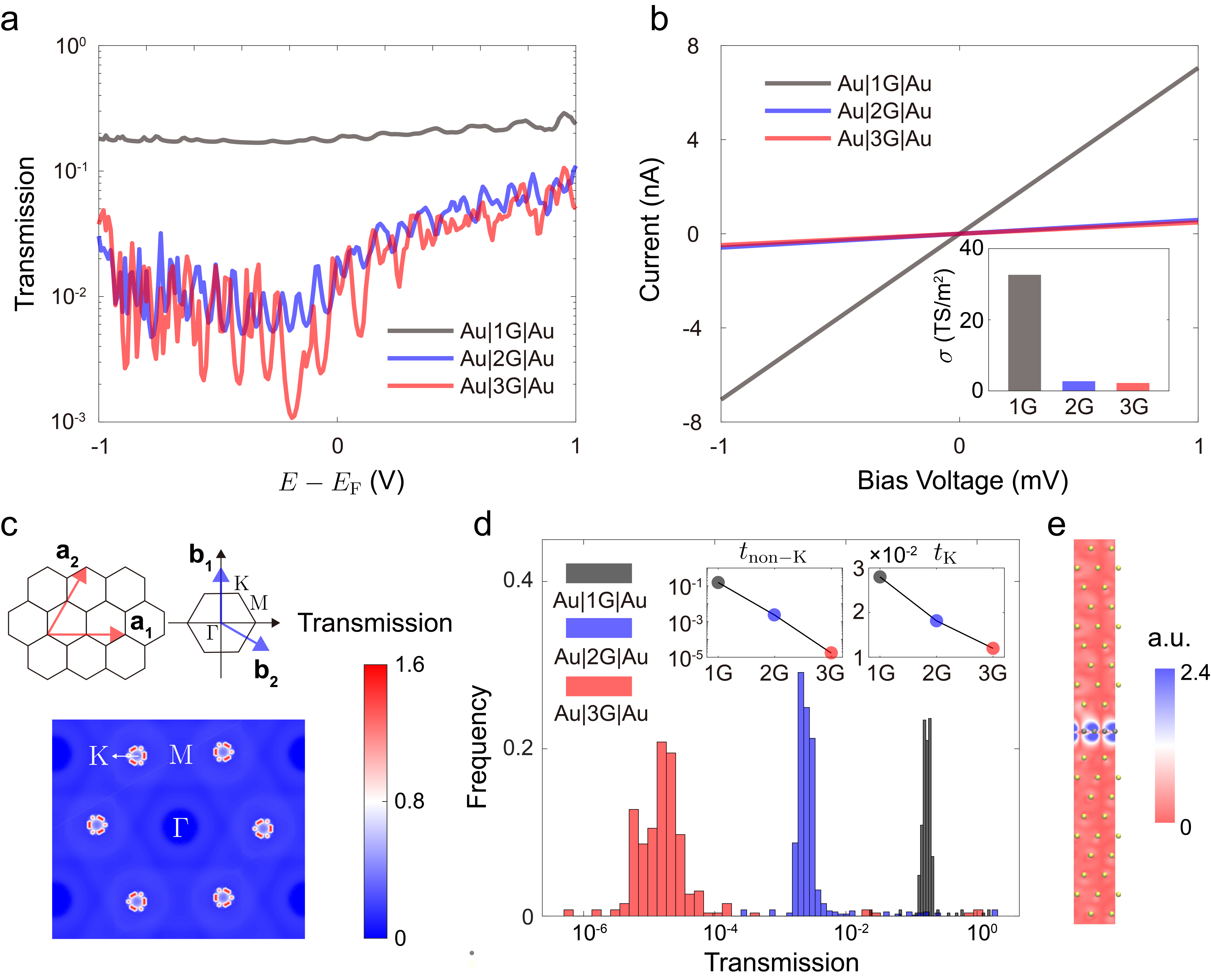}
\caption{
\textbf{Electrical transport across Au$\lvert n$G$\lvert$Au junctions.}
\textbf{a.} Transmission coefficients under a bias ranging from $-1$ to $1$ V.
\textbf{b.} Current-voltage curves in the mV range.
\textbf{c.} ${\bf k}$-resolved transmission coefficients at the Fermi level. The inset shows the Au-C supercell in the lateral directions and its Brillouin zone.
\textbf{d.} Statistical distribution of transmission coefficients at different ${\bf k}_{\parallel}$-points.
\textbf{e.} Transport eigenchannel at the ${\bf k}$-points with the highest transmission coefficient ($1$ a.u. = $1$ $\rm Ha^{-0.5} \rm Bohr^{-1.5}$).
$t_{\rm K}$ and $t_{\rm non-K}$ are the contributions from the K- and non-K channels, respectively.
}
\label{fig2}
\end{figure*}

The ITCs of the Au$\lvert$$n$G$\rvert$Au junctions under different biases ($V_{\rm b}$) are calculated (\zfig{Fig. \ref{fig2}a}). 
The transmissions with $n = 1-3$ at zero bias are $t = 0.183$, $0.0204$ and $0.012$, respectively.
This result indicates that electronic transport through the Au$\lvert$$1$G$\rvert$Au junction is much more strongly scattered compared to that across the Au$\lvert 2$G$\rvert$Au and Au$\lvert 3$G$\rvert$Au contacts.
Corresponding current-voltage ($I$-$V$) data obtained by integrating the transmission over a bias interval suggests an almost linear relationship (\zfig{Fig. \ref{fig2}b.})
Similarly, we find that the interfacial electrical conductance (IEC, $\sigma$ or $G$) across the Au$\lvert$1G$\rvert$Au contact is much larger than that of the $2{\rm G}$ and $3{\rm G}$ junctions.
Specifically, the IEC of junctions with $n = 1-3$, measured from the slopes of $I-V$ curves at low bias, is $32.6$, $2.72$ and $2.23$ TSm$^{-2}$, respectively.
This indicates that the van der Waals interface between graphene layers effectively decouples electronic coupling between the electrodes, whereas the direct Au-C interfaces remain relatively transparent to electrons.

To understand the impact of Au-C and C-C interfaces on electrical conductance, we analyze the ${\bf k}$-resolved transmission, $t\left({\bf k_{\parallel}}\right)$, at the Fermi level (\zfig{Fig. \ref{fig2}c}).
The output suggests that channel-specified transmission peaks at the high-symmetry K points in the Brillouin zone of the Au$\lvert$$n$G$\rvert$Au supercell.
We further decompose $t\left({\bf k_{\parallel}}\right)$ into basal-plane ${\bf k_{\parallel}}$-points (\zfig{Fig. \ref{fig2}d}).
While statistical analysis confirms that K-points within the Brillouin zone are the primary contributors to total channelwise transmission, other regions also play a significant role, as evidenced by the presence of prominent and broad peaks.
Interestingly, the contribution from these non-K points weakens considerably as the number of graphene layers ($n$) in the junction increases.
This observation suggests a tunneling-dominant transport mechanism.

We also extract transmission eigenchannels at K from the transmission matrix ${\bf T}$, which demonstrates the spatially localized nature of the $\pi$ orbitals in graphene layers (\zfig{Fig. \ref{fig2}e}).
Therefore, our findings demonstrate that although electronic coupling between K-states in graphene significantly contributes to electronic transport, the bulk non-K contribution from low-symmetry ${\bf k}$-points reduces the overall conductance.
These findings explain the observed decay of transmission across the interface as $n$ increases, that is, $G \sim e^{-n}$.

\begin{figure*}[htbp]
\centering
\includegraphics[scale = 0.8] {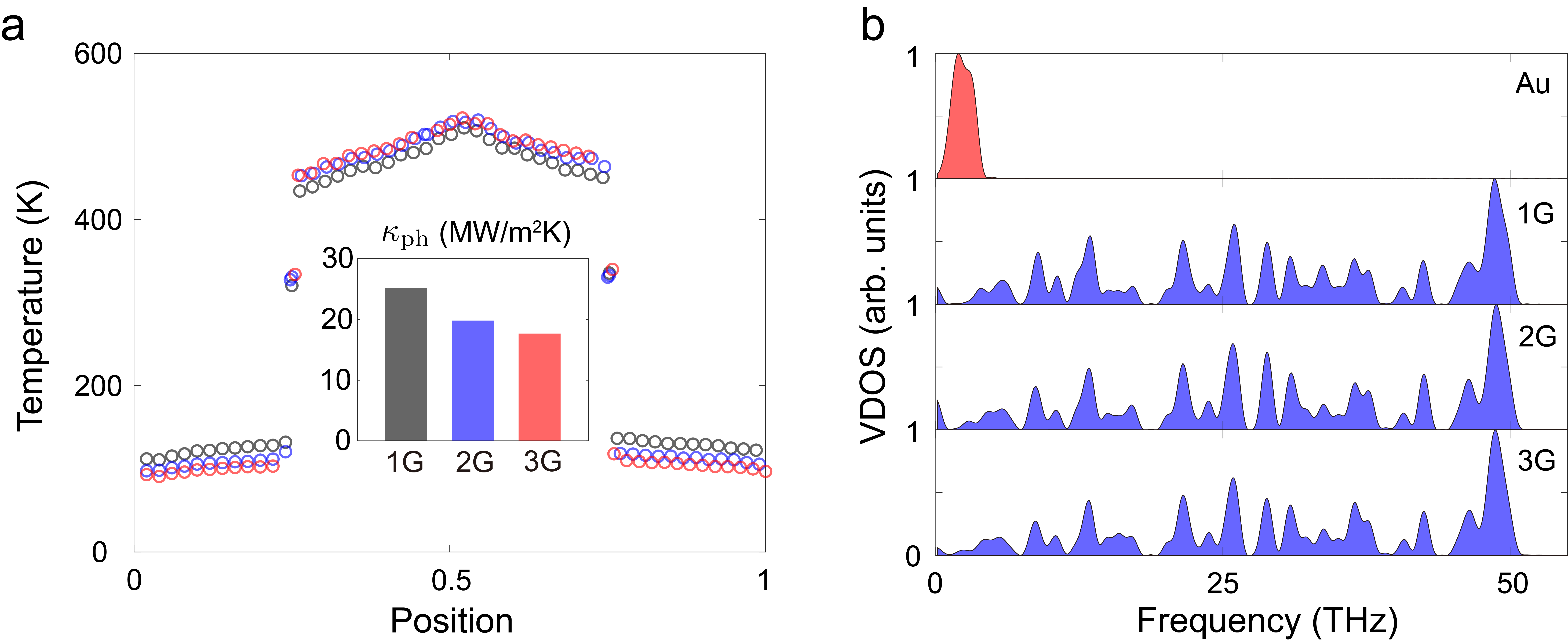}
\caption{\textbf{Phononic thermal transport across Au$\lvert n$G$\lvert$Au junctions.}
\textbf{a.} Temperature profiles and phononic interfacial thermal conductances (ITCs, $\kappa_{\rm ph}$) calculated from NEMD simulations.
\textbf{b.} Vibrational density of states (VDOS) calculated from equilibrium MD simulations for Au atoms and C atoms at the contact.
}
\label{fig3}
\end{figure*}

NEMD simulations model the lattice contribution to ITCs for the Au$\lvert n$G$\lvert$Au junctions.
The temperature profiles demonstrate the thermal barrier at the junction (\zfig{Fig. \ref{fig3}a}).
A closer examination reveals a significant temperature reduction at the Au-C interface compared to the G-G interface, suggesting that the ITC ($\kappa_{\rm ph}$) at the Au-C contact is the limiting factor.
As a result, the phononic thermal conductance of the devices with $n = 1-3$ is close, with values of $\kappa_{\rm ph} = 25.1$, $19.8$ and $17.7$ MWm$^{-2}$K$^{-1}$, respectively.
Our results are quantitatively consistent with the ITCs of metal/graphene interfaces reported in the literature~\cite{wang2016intercalated}.

The thermal coupling between weakly interacting surfaces, such as the models studied in this work, can be assessed by examining the overlap of their vibrational density of states (VDOS)~\cite{wang2016water}.
Examination of the Au$\lvert n$G$\lvert$Au junctions indicates that graphene's characteristics are effectively maintained as $n$ varies from $1$ to $3$, indicating that interfacial coupling does not significantly alter its intrinsic phononic properties.
The observed limited overlap in VDOS at lower frequencies accounts for the low ITC across the Au$\lvert n$G$\lvert$Au junctions, with nearly constant values as $n$ increases.

Using NEGF calculations and NEMD simulations, widely employed to model electronic and phononic thermal transport, we find that $\kappa_{\rm ph}$ remains independent of the number of graphene layers due to the weak coupling between gold and graphene.
However, transmission across Au$\lvert n$G$\lvert$Au, primarily through tunneling, results in a diminishing contribution to the total thermal conductance as $n$ increases.
Based on these findings, we propose an experimental method to determine the Lorenz number at weakly coupled 2D interfaces such as the gold-graphene contact studied here, which could potentially extend to other metal/2D material interfaces.
This approach effectively isolates the electronic contribution from the influence of phonon thermal conductance.

To implement this method, one should measure the electrical and thermal conductance of junctions embedding single-layer (SL) or multi-layer (ML) 2D materials.
The following formula can then be used to calculate the Lorenz number
\begin{eqnarray}
L\left(T\right)T=\frac{\kappa_{\rm SL}-\kappa_{\rm ML}}{\sigma_{\rm SL}},
\end{eqnarray}

\noindent where $\kappa$ and $\sigma$ are the ITC and IEC, respectively.
However, it should be noted that the form of the extended interfacial Wiedemann-Franz Law for metal/2D material interfaces is speculative and requires further evaluation by studying various interfaces and different temperatures. Mathematically, the relationship is expressed as $\kappa/\sigma = L\left({\rm interface}, T\right)T$.

\begin{figure*}[htbp]
\centering
\includegraphics[scale = 0.8] {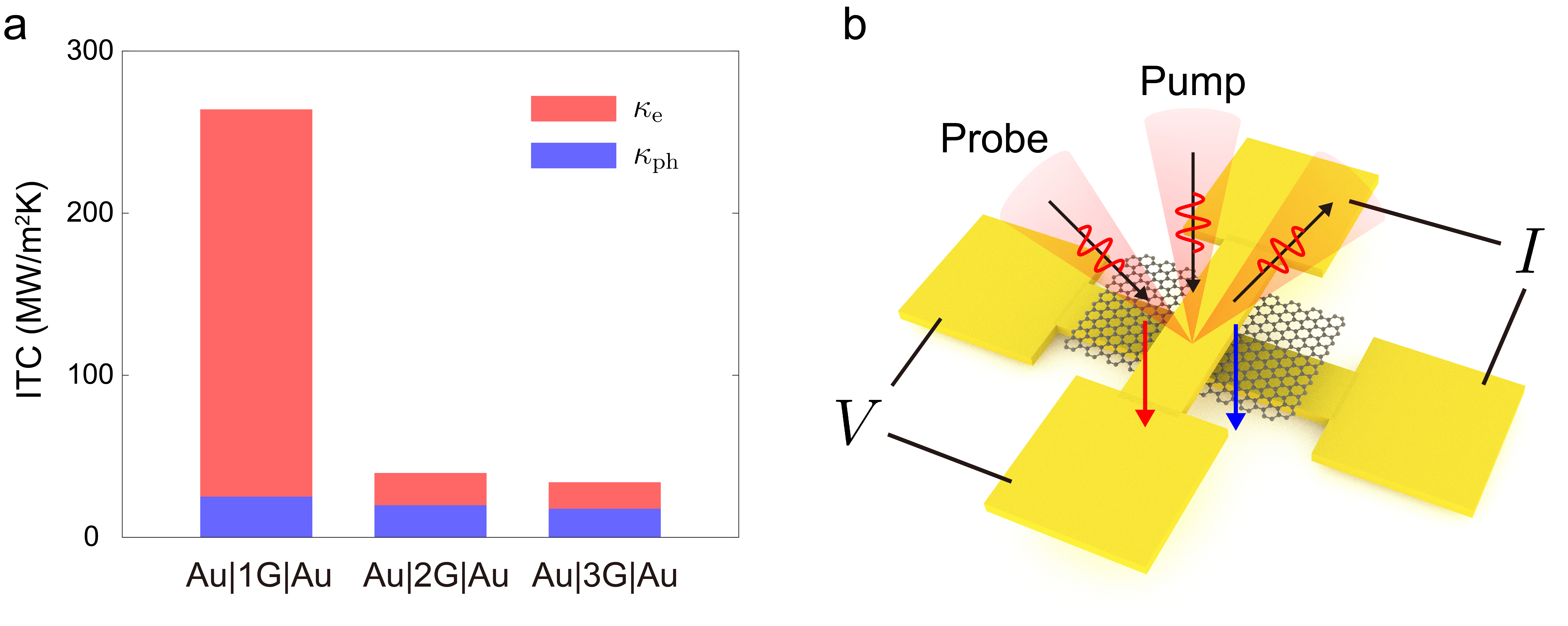}
\caption{\textbf{Proposed setup to disentangle electronic and thermal transport across 2D junctions.}
\textbf{a.} Electronic and phononic ITCs ($\kappa_{\rm e}$, $\kappa_{\rm ph}$) across Au$\lvert n$G$\lvert$Au junctions.
The electronic contribution is calculated by following the Wiedemann-Franz (WF) law with a constant of $L_{0} = 2.44\times 10^{-8}$ W$\Omega \rm{K}^{-2}$ at $T = 300$ K.
\textbf{b.} Schematic diagram of experimental setup via four-terminal and the time-domain thermal
reflection (TDTR) tests.
}
\label{fig4}
\end{figure*}

By simply assuming that the Lorenz number of Au$\lvert n$G$\lvert$Au junctions is a constant and equals to $L_{0} = 2.44\times10^{-8}$ W$\Omega$K$^{-2}$.
The electronic contribution to ITC ($\kappa_{\rm e}$) can be calculated as $\kappa_{\rm e}  =L_{0}G_{0}\tau T/A$.
The results allow us to compare the electronic and phononic contributions to the total thermal conductance (\zfig{Fig. \ref{fig4}a}), the value of which for the junction with a single 2D layer is much higher than that with $n > 1$, where the phononic contribution may dominate.

While the concept is promising, realizing it presents technical challenges. Fortunately, fabrication of devices with single- or few-layer 2D materials like graphene and MoS$_{2}$ is achievable ~\cite{chen2013layer, liu2021transferred}.
This allows for both electrical and thermal conductance measurements within the same vertical setup using the four-terminal method and the time-domain thermal reflection (TDTR) method, as illustrated in \zfig{Fig. \ref{fig4}b}.
Measurements can be performed on devices containing different numbers of 2D material layers and at different temperatures.
However, the surface cleanliness of the 2D materials is crucial, as it significantly impacts interfacial contact conditions and contact resistance, leading to potential inconsistencies in the data.

Using the non-equilibrium Green's function method, we studied how adding graphene layers between gold electrodes affects electrical and thermal conductance.
Our results show that increasing graphene layers significantly reduces electrical conductance due to van der Waals shielding of quantum tunneling.
However, weak bonding between gold and graphene leaves phononic thermal conductance largely unaffected.
The unusual layer dependence allows us to distinguish the electronic contribution to thermal conductance in single-layer graphene devices by subtracting the conductance of multilayer devices.
This theoretical proposal paves the way for theoretical exploration of the extended Wiedemann-Franz Law at low-dimensional material interfaces.

\section*{Acknowledgement}
This study was supported by the National Natural Science Foundation of China through grants 12425201 and 52090032 and the National Key Basic Research Program of China through grant 2022YFA1205400.
L.Z. thanks Dr. Yanlei Wang and Ke Zhou for their help in setting up the simulation models.
The computation was performed on the Explorer 1000 cluster system of the Tsinghua National Laboratory for Information Science and Technology.


\clearpage
\newpage

\section*{Figures and Figure Captions}

\clearpage
\newpage

\clearpage
\newpage

\clearpage
\newpage

\clearpage
\newpage

\bibliography{main_text}

\end{document}